\documentclass[conference]{IEEEtran}
\IEEEoverridecommandlockouts
\usepackage{cite}
\usepackage{amsmath,amssymb,amsfonts}
\usepackage{algorithm}
\usepackage{algorithmic}
\usepackage{hyperref}
\usepackage{cleveref}
\usepackage{array}
\usepackage{graphicx}
\usepackage{subfigure}
\usepackage{float}
\usepackage{graphicx}

\usepackage{textcomp}
\usepackage{xcolor}
\usepackage{hyperref}
\usepackage{pifont}
\usepackage{tabularray}
\def\BibTeX{{\rm B\kern-.05em{\sc i\kern-.025em b}\kern-.08em
    T\kern-.1667em\lower.7ex\hbox{E}\kern-.125emX}}
    
\begin{document}

\title{TempoScale: A Cloud Workloads Prediction Approach Integrating Short-Term and Long-Term Information

\thanks{This work is supported by the National Natural Science Foundation of China (No. 62102408), Shenzhen Industrial Application Projects of undertaking the National key R \& D Program of China (No. CJGJZD20210408091600002), Shenzhen Science and Technology Program (No. RCBS20210609104609044), and Chinese Academy of Sciences President's International Fellowship Initiative (Grant. 2023VTB0005).}
}

\author{\IEEEauthorblockN{Linfeng Wen$^{1, 2}$, Minxian Xu$^{1}$, Adel N. Toosi$^{3}$, Kejiang  Ye$^{1}$}
\IEEEauthorblockA{1. Shenzhen Institute of Advanced Technology, 
Chinese Academy of Sciences, 
Shenzhen, China\\
2. University of Chinese Academy of Sciences, China\\
3. Faculty of Information Technology, Monash University, Australia\\
\{lf.wen, mx.xu\}@siat.ac.cn, adel.n.toosi@monash.edu, kj.ye@siat.ac.cn}}

% \author{\IEEEauthorblockN{Anonymous Authors}}

\maketitle

\begin{abstract}
Cloud native solutions are widely applied in various fields, placing higher demands on the efficient management and utilization of resource platforms. To achieve the efficiency, load forecasting and elastic scaling have become crucial technologies for dynamically adjusting cloud resources to meet user demands and minimizing resource waste. However, existing prediction-based methods lack comprehensive analysis and integration of load characteristics across different time scales. For instance, long-term trend analysis helps reveal long-term changes in load and resource demand, thereby supporting proactive resource allocation over longer periods, while short-term volatility analysis can examine short-term fluctuations in load and resource demand, providing support for real-time scheduling and rapid response. In response to this, our research introduces TempoScale, which aims to enhance the comprehensive understanding of temporal variations in cloud workloads, enabling more intelligent and adaptive decision-making for elastic scaling. 

TempoScale utilizes the Complete Ensemble Empirical Mode Decomposition with Adaptive Noise algorithm to decompose time-series load data into multiple Intrinsic Mode Functions (IMF) and a Residual Component (RC). First, we integrate the IMF, which represents both long-term trends and short-term fluctuations, into the time series prediction model to obtain intermediate results. Then, these intermediate results, along with the RC, are transferred into a fully connected layer to obtain the final result. Finally, this result is fed into the resource management system based on Kubernetes for resource scaling. Our proposed approach can reduce the Mean Square Error by 5.80\% to 30.43\% compared to the baselines, and reduce the average response time by 5.58\% to 31.15\%. The results demonstrate the effectiveness of our proposed method in reducing violations of service-level objectives and providing better performance in terms of resource utilization.
\end{abstract}


\section{Introduction}
With the rise of cloud native and microservices architecture, containerization technology has emerged as a crucial innovation, profoundly altering the landscape of software development and deployment \cite{b2}. Among these technologies, Kubernetes (K8s), as an open-source container orchestration platform, has provided a robust framework for automating the deployment, scaling, and operation of application containers, thereby significantly enhancing efficiency \cite{b5}. Nowadays, K8s has been widely adopted by mainstream companies, including Amazon, Google, and Microsoft.

However, as more and more enterprises adopt containerized microservices architectures, and application scenarios continue to evolve, limitations of the reactive strategy employed by the default resource scheduler in K8s have become apparent \cite{10255032}. For instance, under highly variable workloads, its applicability is limited, leading to resource waste and a decrease in service quality. Consequently, some enterprises are gradually shifting towards predictive scaling methods. Predictive scaling not only focuses on current loads but also involves analyzing historical data, trends, and predictive models to forecast future loads and make adjustments based on this analysis. The advantage of this approach lies in proactively allocating resources, avoiding the need for reactive measures when loads increase suddenly. Currently, elastic scaling based on load forecasting has become a crucial technology for effectively adjusting cloud resources in dynamic environments to meet user needs and minimize resource waste.

Applying a prediction-based strategy in a production environment still faces several challenges: Firstly, the dynamics and uncertainty of system and network workloads render traditional static models inadequate when confronted with complex and rapidly changing workloads \cite{1,10254982}. Additionally, the diversity and heterogeneity of real-world workloads further complicate predictions, making it difficult for a singular approach to adapt to various scenarios and environments \cite{b8}. Lastly, existing methods lack the ability to extract features across different time sequences, resulting in a lack of comprehensive understanding of the workloads. The inherent dynamism of these clusters, the variability of workloads, and the inherent limitations of the method itself, presents challenges that demand innovative solutions \cite{10077424}. Therefore, this research attempts to address these challenges by exploring and extracting features such as CPU utilization at different time scales, proposing a comprehensive load prediction method that integrates both long-term and short-term time series information.

In terms of long-term load forecasting, we have adopted advanced time series and machine learning methods to process and extract features from historical load data, establishing an accurate long-term load forecasting model. Simultaneously, to better capture instantaneous fluctuations in the system, we  construct an effective short-term load prediction model through meticulous data sampling and feature extraction. To integrate the information from long-term and short-term load forecasting, we utilize the Complete Ensemble Empirical Mode Decomposition with Adaptive Noise (CEEMDAN) \cite{b9} to decompose the load time series data. This step generates multiple Intrinsic Mode Functions (IMFs) and a Residual Component (RC), representing load changes at different time scales. By integrating these IMFs and RC through a fully connected layer, we obtain more comprehensive and detailed load prediction results, enhancing the system's robustness.

Finally, by developing elastic scaling decision rules based on integrated load prediction, we make cloud computing platforms more adaptable and able to flexibly adjust resource allocation according to real-time load conditions. This method not only improves the performance and stability of the system but also effectively reduces resource costs and promotes the sustainable development of cloud computing across various scenarios.

The \textbf{  main contributions} of this work are:
	
\begin{itemize} 

	\item [$ \bullet $] We utilize CEEMDAN to divide the load into three modes: IMF of long-term trend, IMF of short-term fluctuation, and RC. This approach helps capture information across both long-term and short-term time series by separating features.
	
	\item [$ \bullet $]We present TempoScale, a cloud workload prediction method that integrates short-term fluctuations and long-term trends. This approach enhances a comprehensive understanding of temporal variations in loads, enabling more intelligent and adaptive decision-making for elastic scaling.
		
	\item [$ \bullet $] We evaluate the effectiveness and availability of several baselines through realistic traces and testbed. The results demonstrate the effectiveness of our proposed method in reducing violations of service level objectives (SLO) and improving performance in terms of resource utilization.
		
\end{itemize}

% The rest of the paper is organized as follows: Section~\ref{sec:Motivation} provides a detailed exploration of the motivation and feasibility by combining short-term and long-term information. Section~\ref{sec:Related Work} discusses related work on prediction-based elastic scaling for cloud applications.  Section~\ref{sec:TempoScale} presents the time series forecasting algorithm of TempoScale. Section~\ref{sec:Performance} demonstrates the performance evaluations of the proposed approach. Finally, Section~\ref{sec:Conclusions} concludes the paper and highlights promising future directions.

\section{Motivation and Feasibility}\label{sec:Motivation}
\begin{figure}
	\centering
	\subfigure[Long-Term Prediction that Shows Trend.]{\label{fig:LONG}
		\includegraphics[width=0.49\textwidth]{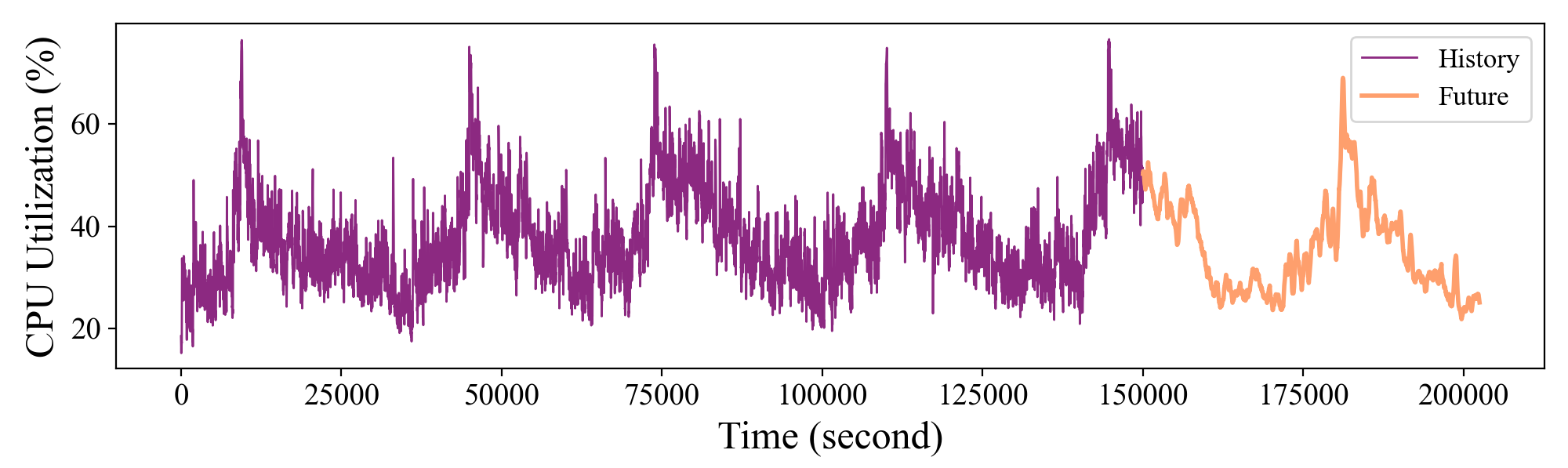}}
	\subfigure[Short-Term Prediction that Shows Instantaneous Fluctuations.]{\label{fig:SHORT}
		\includegraphics[width=0.49\textwidth]{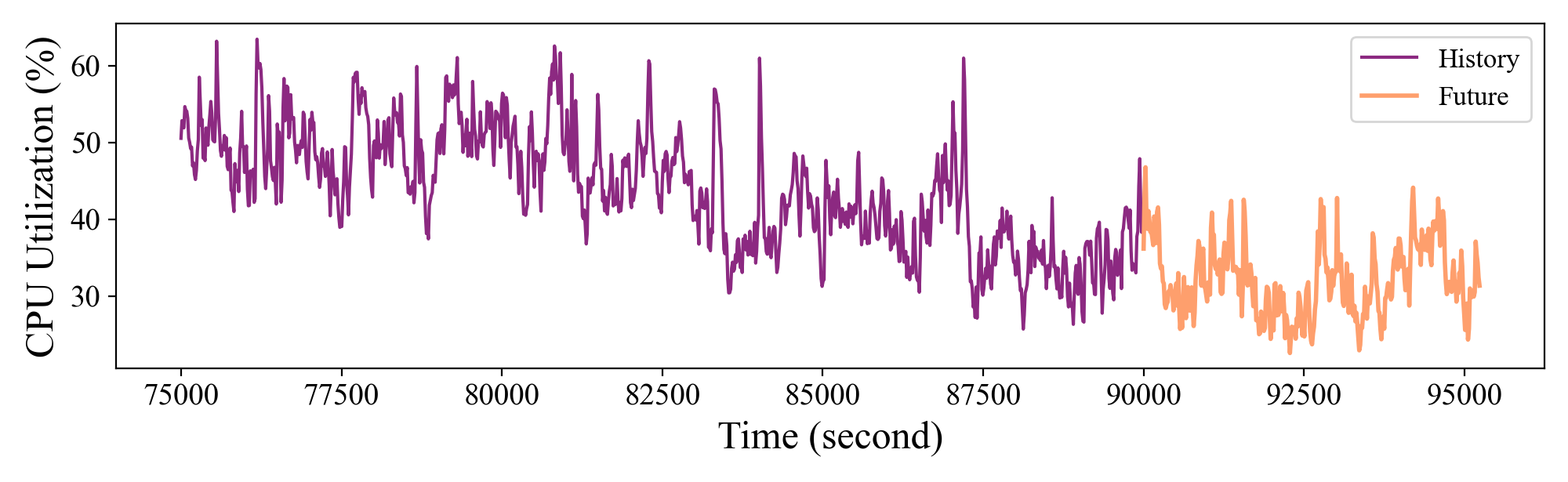}}
	\caption{The Focus Differs Between Long-Term and Short-Term Predictions in Time Series Forecasting.}
	\label{fig:long-short}
\end{figure}

Existing load forecasting methods often focus on either the short-term fluctuations of the load or the long-term trends, lacking an integrated model that combines both short-term volatility and long-term trends. As shown in Fig.~\ref{fig:long-short}, utilizing long-term trend analysis can reveal prolonged changes in load and resource requirements, such as weekly or daily periodic variations, supporting the proactive allocation of resources over an extended period. On the other hand, employing short-term volatility analysis allows for the examination of short-term fluctuations in load and resource demands, including peak and off-peak loads, as well as sudden spikes, supporting real-time scheduling and rapid response.

The integration of short-term volatility and long-term trends in the study of load variations in large-scale systems can further enhance the precision of models characterizing resource requirements. This, in turn, provides robust support for the performance optimization of elastic scheduling.

To illustrate the limitations of approaches focusing solely on either short-term or long-term aspects, we conducted preliminary  forecasting experiments to calculate the Mean Square Error (MSE) using experimental data from the Alibaba Cluster\footnote[1]{https://github.com/alibaba/clusterdata/tree/master/cluster-trace-v2018}, which provides real production cluster traces. We selected two representative models for long-term and short-term time series prediction, namely Informer \cite{11} and efficient supervised learning-based Deep Neural Network (esDNN \cite{2}). Informer, based on the Transformer architecture, demonstrating significant improvements in long-term predictive performance compared to the original Transformer. On the other hand, esDNN, based on Gated Recurrent Unit (GRU) \cite{b12}, is an algorithm used for short-term cloud load prediction. It adapts to workload variations by updating GRU control gates, overcoming limitations such as gradient vanishing and exploding. For different forecast horizons, we employed esDNN and Informer respectively for time series forecasting of loads. Each experiment was repeated 10 times, and the experimental results are presented in Table~\ref{tab:esDNN and Informer}. The collected results have all been subjected to inverse  normalization.

\begin{table}
\caption{Performance Comparison of Short-Term (esDNN) and Long-Term (Informer) Prediction.}
\label{tab:esDNN and Informer}
\centering
\resizebox{\linewidth}{!}{%
\begin{tblr}{
  width = \linewidth,
  colspec = {Q[215]Q[206]Q[229]Q[283]},
  cells = {c},
  hline{1,9} = {-}{0.08em},
  hline{2} = {-}{0.05em},
}
History:Future & esDNN (MSE) & Informer (MSE) & {Informer is X\% \\better than esDNN}\\
3:1 & \textbf{3.65} & 4.12 & -12.88\%\\
6:2 & \textbf{5.18} & 5.70 & -10.04\%\\
12:4 & \textbf{8.07} & 8.68 & -7.56\%\\
24:8 & 13.06 & \textbf{13.05} & 0.08\%\\
48:16 & 19.61 & \textbf{18.87} & 3.77\%\\
96:32 & 28.11 & \textbf{26.69} & 5.05\%\\
192:64 & 37.53 & \textbf{33.71} & 10.18\%
\end{tblr}
}
\end{table}

\begin{figure}
	\centerline{\includegraphics[width=\linewidth]{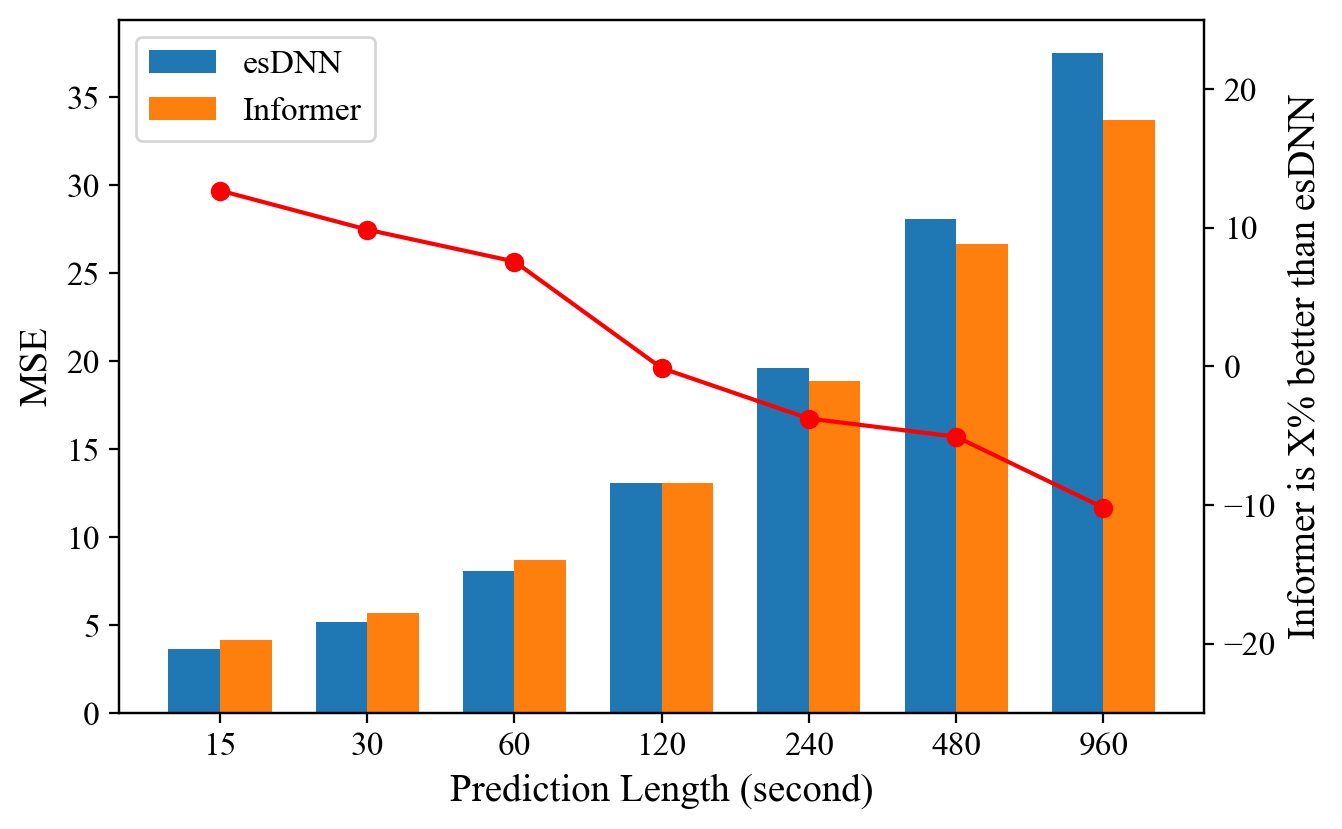}}
	\caption{Performance Comparison of Short-Term (esDNN) and Long-Term (Informer) Prediction.}\label{fig:esDNN_Informer}
\end{figure}

Fig.~\ref{fig:esDNN_Informer} illustrates that in short-term time series prediction, particularly when the predicted time series length is less than 8 points (each point having a 30-second interval), Informer performs noticeably worse than esDNN. On the other hand, in long-term time series prediction, when the predicted time series length exceeds 8 points, Informer outperforms esDNN significantly. Moreover, as the time series length increases or decreases, the performance gap becomes more significant. This insight suggests the possibility of performance optimization through the development of an integrated algorithm that combines long-term and short-term time series prediction.

\section{Related Work}\label{sec:Related Work}
Many researchers have extensively investigated workload forecasting, and these studies can be categorized into three classes: 1) machine learning-based models, 2) neural network-based models, and 3) attention mechanism-based models.

\subsection{Traditional Machine learning-based models}
Numerous studies have been dedicated to leveraging traditional machine learning models to enhance the accuracy and efficiency of cloud workload prediction. 

% Research in this field encompasses various approaches to adapt to the ever-changing cloud computing environment, such as Support Vector Machines, Decision Trees, Exponential Smoothing, Autoregressive Integrated Moving Average (ARIMA).

Prassanna et al. \cite{6} proposed a new virtual machine consolidation technique, NMT-FOLS, which employs an Adaptive Regressive Holt-Winters Workload Predictor to identify the workload state and utilizes the prediction results to allocate user-requested tasks to the optimal VM. Xie et al. \cite{7} proposed a hybrid model of ARIMA and triple exponential smoothing, which accurately predicts both linear and nonlinear relationships in the container resource load sequence. The weighting values of the two 
different models are chosen based on the sum of squares of their prediction errors over a period of time. Biswas et al. \cite{8} proposed a new Linear Regression model to predict future CPU utilization, which is described by a straight line and a mean point. The proposed algorithm reduces energy consumption and SLA violation rates in cloud data centers. Kholidy \cite{9} developed a novel Swarm Intelligence-Based Prediction Approach, which utilizes Particle Swarm Optimization to select optimal features from the dataset and estimate parameters for the prediction algorithms. Righi et al. \cite{10} proposed a proactive elasticity model named Proliot. The contribution of Proliot lies in its utilization of a mathematical formalism employing ARIMA and Weighted Moving Average for predicting the behavior of Internet of Things load, enabling anticipation of scaling operations.

\subsection{Neural network-based models}
Traditional machine learning methods perform poorly when dealing with complex features of cloud workloads. To more effectively capture these features, researchers have turned to more advanced learning methods, with deep learning being particularly noteworthy \cite{b13}, especially excelling in handling large-scale, high-dimensional data, and nonlinear relationships.
% Deep learning methods are favored for their superior ability to learn complex representations, 

Dogani et al. \cite{1} proposed an innovative approach utilizing Bidirectional GRU and Discrete Wavelet Transformation to enhance the accuracy of host workload prediction. Xu et al. \cite{2} proposed an esDNN algorithm for cloud workload prediction, which adapts to workload variations by updating the gates of the GRU, overcoming the limitations of gradient disappearance and explosion. Ruan et al. \cite{4} proposed a deep learning-based workload prediction method named CrystalLP that utilize Long Short-Term Memory (LSTM) networks. Ouhame et al. \cite{5} proposed a Convolutional Neural Network (CNN) and LSTM model for predicting multivariate workloads. It first analyzes the input data using the vector autoregression method to filter the linear correlations among multivariate data. Then, it calculates the residual data and inputs it into the CNN layer to extract complex features of each virtual machine usage component.

\subsection{Attention mechanism-based models}
Attention mechanism is a new research field base on neural networks in recent years, achieving significant success \cite{b11}. The core mechanism of attentional mechanisms involves focusing resources on key components of time series input while filtering out irrelevant information \cite{11}. Therefore, attention mechanisms can enhance the model's capability to capture the dynamic changes in time series data, thereby enhancing the performance of time series analysis and prediction tasks.
% Currently, researchers are beginning to introduce attention mechanisms into the domain of time series \cite{11}. improving its modeling of long-term dependencies within sequences and 

Zhou et al. \cite{11} designed an efficient Transformer model for Long Sequence Time Series Forecasting named Informer. It employs the ProbSparse self-attention mechanism, self-attention distilling by halving cascading layer input, and a generative-style decoder, significantly improving the inference speed for long sequence predictions. Zerveas et al. \cite{12} introduced a novel framework for multivariate time series representation learning based on the Transformer encoder architecture. The framework incorporates an unsupervised pre-training scheme, which offers substantial performance benefits over fully supervised learning in downstream tasks. Wang et al. \cite{14} proposed a novel method for time series prediction, leveraging the Transformer with a multiscale CNN. It consists of multiscale extraction and multidimensional fusion  frameworks. Wu et al. \cite{15} designed Autoformer as a novel decomposition architecture with an Auto-Correlation mechanism, breaking the preprocessing convention of time series decomposition and transforming it into a fundamental building block of deep models. This design empowers Autoformer with progressive decomposition capabilities for complex time series.

% Wang et al. \cite{13} proposed R-Transformer, which combines the advantages of both Recurrent Neural Network (RNN) and the multi-head attention mechanism while avoiding their respective drawbacks. The proposed model can effectively capture both local structures and global long-term dependencies in sequences without the use of any position embeddings.

% \usepackage{tabularray}
\begin{table}
\centering
\caption{Related Work.}
\label{tab:related work}
\resizebox{\linewidth}{!}{%
\begin{tblr}{
  column{1} = {c},
  column{3} = {c},
  column{4} = {c},
  column{5} = {c},
  column{7} = {c},
  column{8} = {c},
  column{10} = {c},
  cell{1}{1} = {r=2}{},
  cell{1}{2} = {r=2}{},
  cell{1}{3} = {c=3}{},
  cell{1}{7} = {c=2}{},
  cell{1}{10} = {r=2}{},
  hline{1,17} = {-}{},
  hline{2} = {3-5,7-8}{},
  hline{3} = {-}{0.05em},
}
Work &  & Types of Prediction Methods &  &  &  & Prediction Length &  &  & {Resource\\Scaling\\in Cloud }\\
 &  & {Machine\\Learning} & {Neural\\Network} & {Attention\\Based} &  & Single & Multiple &  & \\
Dogani et al. \cite{1} &  &  & \checkmark & \checkmark &  & \checkmark &  &  & \\
Zhou et al. \cite{11} &  &  &  & \checkmark &  &  & \checkmark &  & \\
Xu et al. \cite{2} &  &  & \checkmark &  &  & \checkmark &  &  & \checkmark\\
Prassanna al. \cite{6} &  & \checkmark &  &  &  & \checkmark &  &  & \checkmark\\
Xie et al. \cite{7} &  & \checkmark &  &  &  &  & \checkmark &  & \checkmark\\
Biswas et al. \cite{8} &  & \checkmark &  &  &  & \checkmark &  &  & \checkmark\\
Kholidy \cite{9} &  & \checkmark &  &  &  &  & \checkmark &  & \\
Righi et al. \cite{10} &  & \checkmark &  &  &  & \checkmark &  &  & \\
Ruan et al. \cite{4} &  &  & \checkmark &  &  & \checkmark &  &  & \\
Ouhame et al. \cite{5} &  &  & \checkmark &  &  &  & \checkmark &  & \\
Zerveas et al. \cite{12} &  &  &  & \checkmark &  &  & \checkmark &  & \\
Wang et al. \cite{14} &  &  &  & \checkmark &  &  & \checkmark &  & \\
Wu et al. \cite{15} &  &  &  & \checkmark &  &  & \checkmark &  & \\
This paper &  &  & \checkmark & \checkmark &  &  & \checkmark &  & \checkmark
\end{tblr}
}
\end{table}

\subsection{Critical analysis}

We summarize and compare the related work in Table~\ref{tab:related work}. Models based on traditional machine learning are mostly effective for workloads with clear patterns, but the high variability and non-linearity of modern cloud workloads make these models less effective \cite{10254959}. Neural network-based models may face issues like gradient vanishing or exploding when dealing with long sequences, especially in tasks that require considering long-term dependencies. This can make it challenging for the model to capture and learn effective information over extended time intervals. Transformer-based models, leveraging attention mechanisms, excel in long-time sequence prediction tasks, significantly improving performance. However, they also come with drawbacks, such as larger parameter sizes, complex tuning processes, and higher resource costs, leading to increased usage expenses.

Therefore, this study proposes a predictive algorithm that integrates both long-term and short-term temporal features, enabling better capture of dependencies in time series data. Utilizing long-term trend analysis to reveal the extended variations in load and resource demands supports proactive resource allocation over more extended periods. Additionally, employing short-term volatility analysis examines the short-term variations in load and resource demands, facilitating real-time scheduling and rapid responsiveness. The main difference between our work and others is that we focus more on studying the characteristics of time series data across different dimensions. We employ different types of models to process them, enabling us to leverage strengths effectively.

\section{TempoScale: A Resource Scheduler Integrating Short-Term and Long-Term Information}\label{sec:TempoScale}

\begin{figure}[H]
	\centerline{\includegraphics[width=\linewidth]{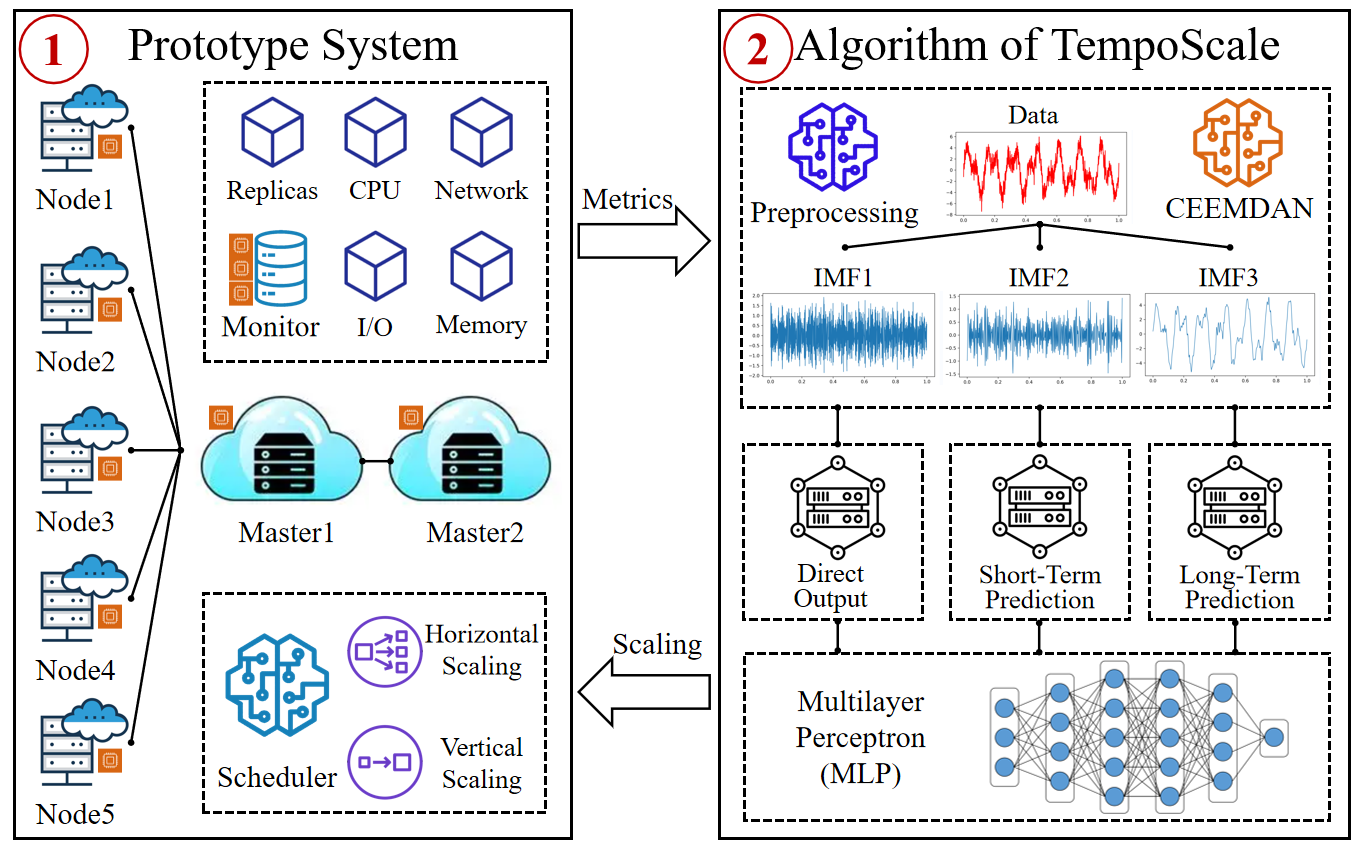}}
	\caption{The Architecture of TempoScale.}\label{fig:architecture}
\end{figure}

In order to address the inherent dynamics of clusters and the variability of workloads, we propose an innovative solution in this work called TempoScale. The architecture of TempoScale is illustrated in Fig.~\ref{fig:architecture}, the module \scalebox{1}{\ding{172}} represents a server cluster, we have implemented a prototype system and deployed a resource scheduler and a resource monitor\footnote[2]{https://prometheus.io/}$^,$\footnote[3]{https://github.com/kubernetes-sigs/metrics-server}, enabling real-time monitoring and resource control of the server cluster, the module \scalebox{1}{\ding{173}} illustrates the TempoScale algorithm, which involves three steps: 1) preprocessing of data and decomposition of IMFs, 2) processing intermediate results using different models, and 3) obtaining the final results through a Multilayer Perceptron (MLP). In the following sections, we will focus on providing a detailed description of these steps.

\subsection{Preprocessing of data and decomposition of IMFs}
TempoScale processes raw data exported from the monitoring system of a cloud cluster ,the monitoring system of the cloud cluster records major resource usage such as CPU, memory, disk, and network. Among these resources, CPU is considered as the most crucial and dominant resource in the computer system, and we primarily focus on the usage of CPU. Initially, TempoScale removes rows containing empty and anomalous data as they can negatively impact predictive data. Subsequently, TempoScale calculates the average value for each parameter with the same timestamp, represented as a time-ordered sequence $X(x_1, x_2, ..., x_t)$ with constant time intervals. Then, it normalizes the data to enhance the model's convergence speed and prediction accuracy. TempoScale utilizes Z-score for this purpose. Z-score normalization assumes that the data approximates a normal distribution. This normalization method helps eliminate scale differences between different features. $Z$ is calculated using Eq.~(\ref{e1}):
\begin{equation}
    \label{e1}
    Z = \frac{(X - \mu)}{\sigma} = \frac{\displaystyle X - \frac{1}{n} \sum_{i=1}^{n} X_i}{\displaystyle \sqrt{\frac{1}{n} \sum_{i=1}^{n} (X_i - \frac{1}{n} \sum_{i=1}^{n} X_i)^2}},
\end{equation}
where \( X \) represents the original data, \( n \) is the number of data points, \( \mu \) is the mean of the original dataset, \( \sigma \) is the standard deviation of the dataset, and \( Z \) is the standardized data value.

\begin{figure}[H]
	\centerline{\includegraphics[width=\linewidth]{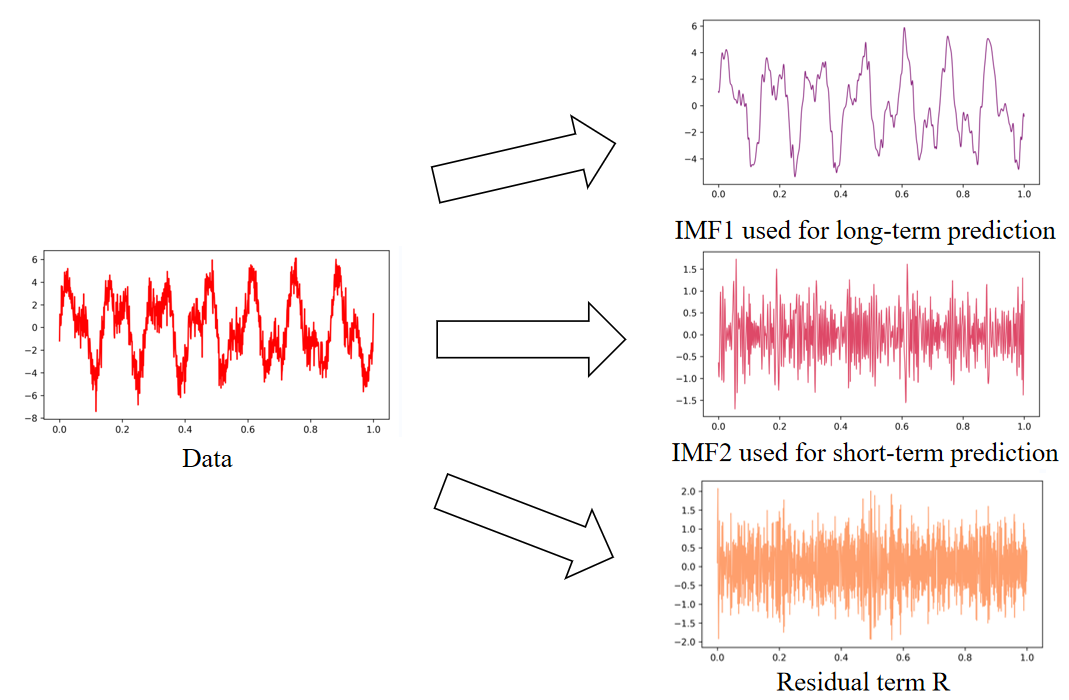}}
	\caption{Modal Decomposition for Feature Extraction.}\label{fig:ceemdan}
\end{figure}

\begin{algorithm}
\caption{Performing Long-Term and Short-Term Feature Segmentation on the Data Within TempoScale.}
\label{alg:ceemdan}
\begin{algorithmic}[1]
    \REQUIRE Signal $x(t)$
    \ENSURE Set of IMFs $\{c_i(t)\}$ and RC $r(t)$

    \STATE Initialize parameters:
    \STATE $N$ - Number of ensemble trials
    \STATE $M$ - Number of sifting iterations
    \STATE $T$ - Signal length
    \STATE $t$ - Time index
    \STATE $c_i(t)$ - Initial IMF estimate
    \STATE $r(t)$ - RC
    \STATE $\alpha$ - Sifting parameter
    \STATE \textbf{Ensemble Generation:}
    \FOR{$i = 1$ to $N$}
        \STATE Generate white noise series $w_i(t)$
        \STATE Add white noise to the signal: $y_i(t) = x(t) + w_i(t)$
        \STATE Perform EMD on $y_i(t)$ to obtain IMF set: $\{c_{i,1}(t), c_{i,2}(t)\}$
    \ENDFOR
    \STATE \textbf{Ensemble Mean:}
    \FOR{$k = 1$ to $2$}
        \STATE Compute ensemble mean of each IMF: $\bar{c}_k(t) = \frac{1}{N} \sum_{i=1}^{N} c_{i,k}(t)$
    \ENDFOR
    \STATE \textbf{Adaptive Noise:}
    \FOR{$k = 1$ to $2$}
        \STATE Compute adaptive noise for each IMF: $a_k(t) = \sqrt{\frac{1}{N} \sum_{i=1}^{N} (c_{i,k}(t) - \bar{c}_k(t))^2}$
    \ENDFOR
    \STATE \textbf{Sifting Process:}
    \FOR{$m = 1$ to $M$}
        \STATE Extract the RC: $r(t) = x(t) - \sum_{k=1}^{2} c_k(t)$
        \FOR{$k = 1$ to $2$}
            \STATE Sift the IMF: $\tilde{c}_k(t) = c_k(t) + \alpha a_k(t)$
        \ENDFOR
    \ENDFOR
\end{algorithmic}
\end{algorithm}

After preprocessing the data, it is necessary to separately extract the long-term and short-term features of the time-series data for subsequent load forecasting. As shown in Fig.~\ref{fig:ceemdan}, modal decomposition decomposes complex signals into IMFs, enhancing the clarity of signal analysis and demonstrating excellent localization in the time-frequency domain. This facilitates a more accurate exploration of the local characteristics of signals. TempoScale leverages the inherent advantages of modal decomposition in time-series feature extraction, which naturally divides preprocessed data into two IMFs and a RC. The two IMFs represent the long-term and short-term features of the time-series data, respectively, and are then used as inputs for subsequent models. As illustrated in Algorithm~\ref{alg:ceemdan}, the algorithm mainly consists of five steps: \textbf{Initializing Parameters}, \textbf{Ensemble Generation}, \textbf{Ensemble Mean}, \textbf{Adaptive Noise}, and \textbf{Sifting Process}.  First, parameters are initialized to set up various necessary parameters for the algorithm (lines 1-7). Then, multiple ensemble trial instances are generated by Empirical Mode Decomposition (EMD) to evaluate the algorithm's robustness under different data variations (lines 8-13). Next, the ensemble means and adaptive noise for each IMF are computed to enhance the accuracy of extracting the actual signal features (lines 14-21). Finally, the sifting process iteratively extracts IMFs and separates the residual signal, achieving the decomposition of the original signal (lines 22-28).

\subsection{Processing intermediate results using different model}
After modal decomposition, two IMFs representing long-term trends and short-term fluctuations will be fed into the prediction module composed of a model based on Transformer and GRU architectures (feasibility has been validated in Section~\ref{sec:Motivation}).

\subsubsection{Short-term prediction model}
The GRU is a type of RNN architecture designed for capturing dependencies and patterns in time series data. The gating mechanisms in GRU allow the network to selectively update and memorize information in the hidden state, enabling it to focus on relevant information while avoiding the long-term dependency issues that can lead to vanishing or exploding gradients. This makes GRU effective at capturing short-term patterns in time series data.

% \begin{algorithm}
%     \caption{The Compilation of the Short-Term Prediction Model in TempoScale.}
%     \label{alg:esDNN}
%     \begin{algorithmic}[1]
%         \REQUIRE 1D Convolutional layer, GRU layer, Dense layer, Dropout layer
%         \ENSURE Compiled model
%         \STATE Create Sequential model
%         \STATE Add 1D Convolutional layer: \(\text{{Conv1D}}(32, 3, 1, \text{{relu}})\)
%         \STATE Add GRU layer: \(\text{{GRU}}(32, \text{{input\_shape}}=(3, 1)\)
%         \STATE Add GRU layer: \(\text{{GRU}}(16, \text{{input\_shape}}=(3, 1))\)
%         \STATE Add Dense layer with ReLU activation: \(\text{{Dense}}(16, \text{{relu}})\)
%         \STATE Add Dropout layer: \(\text{{Dropout}}(0.2)\)
%         \STATE Add Dense layer: \(\text{{Dense}}(1)\)
%     \end{algorithmic}
% \end{algorithm}

Assuming at time step $t$, given input $x_t$, the previous hidden state $h_{t-1}$, and the parameters $W$ of the GRU, we can compute the update gate $z_t$, the reset gate $r_t$, and the candidate hidden state $\tilde{h}_t$ at the current time step \cite{b12}:

\begin{equation}
z_t = \sigma(W_z \cdot [h_{t-1}, x_t]),
\end{equation}
\begin{equation}
r_t = \sigma(W_r \cdot [h_{t-1}, x_t]),
\end{equation}
\begin{equation}
\tilde{h}_t = \tanh(W_h \cdot [r_t \odot h_{t-1}, x_t]),
\end{equation}
where $\sigma$ is the sigmoid function, $\odot$ denotes element-wise multiplication, $[\cdot]$ indicates matrix multiplication, and \text{tanh} represents the hyperbolic tangent function, which is a type of nonlinear activation function:
\begin{equation}
\text{tanh}(x) = \frac{e^x - e^{-x}}{e^x + e^{-x}}.
\end{equation}

Then, based on the update gate $z_t$ and the candidate hidden state $\tilde{h}_t$, we can compute the current hidden state $h_t$ as follows:
\begin{equation}
h_t = (1 - z_t) \odot h_{t-1} + z_t \odot \tilde{h}_t,
\end{equation}
where the update gate $z_t$ controls the balance between the past hidden state $h_{t-1}$ and the new candidate hidden state $\tilde{h}_t$. If $z_t$ is close to 1, the model retains more of the past information; if $z_t$ is close to 0, the model relies more on the new information. The reset gate $r_t$ controls the influence of the past hidden state in computing the candidate hidden state $\tilde{h}_t$.

Through this gating mechanism, GRU effectively captures long-term dependencies and performs well in handling sequential data. Therefore, we adopt the algorithm based on GRU \cite{2} as our short-term prediction model in TempoScale.

\begin{figure}[H]
	\centerline{\includegraphics[width=\linewidth]{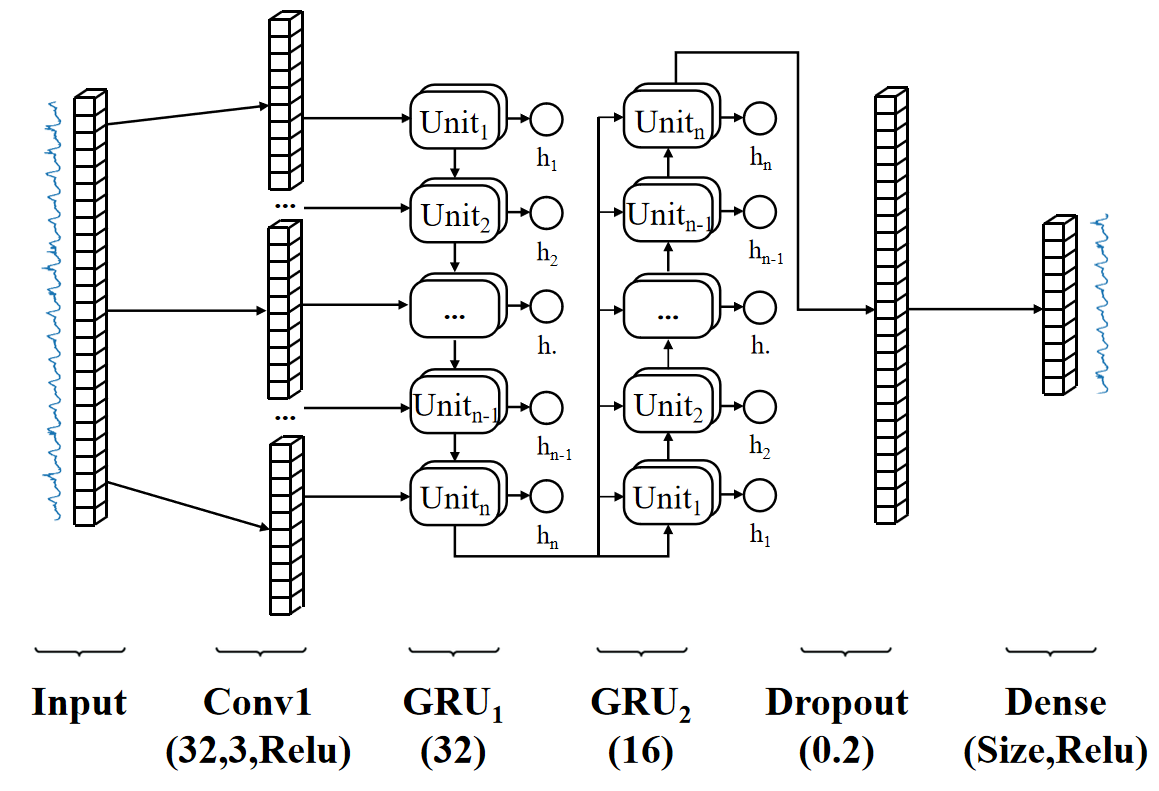}}
	\caption{The Network Structure of Short-term Prediction Model in TempoScale.}\label{fig:esDNN}
\end{figure}

The short-term forecasting algorithm used in TempoScale is depicted in Fig.~\ref{fig:esDNN}. First, the data is fed into a 1D CNN, which can extract features from time series data and model the short-term correlations between the time series data and subsequent trends \cite{b16}. The convolved data is then input into a two-layer GRU network, and finally, the activation function ReLU, regularization, and dense layers are applied to generate the final output. This comprehensive approach ensures a refined and well-optimized prediction based on the extracted features and short-term correlations captured during the earlier stages of processing.

\subsubsection{Long-term prediction model}
Methods based on attention mechanisms typically perform exceptionally well in addressing long time series prediction problems because attention mechanisms enable the model to better focus on different parts of the time series, thereby capturing long-term dependencies more effectively. In self-attention mechanisms, each element in a sequence interacts with every other element to compute weights. Specifically, for each attention head $i$, the attention score matrix $\text{Attention}_i$ can be calculated using Eq.~(\ref{eq:attention_i}):
\begin{equation}
\text{Attention}_i = \text{softmax}\left(\frac{\mathbf{Q}_i \cdot \mathbf{K}_i^T}{\sqrt{d_k}}\right) \cdot \mathbf{V}_i.
\label{eq:attention_i}
\end{equation}

Here, $\mathbf{Q}_i$, $\mathbf{K}_i$, and $\mathbf{V}_i$ are the query, key, and value matrices for the $i$-th head, and $d_k$ is the dimension of the key vectors. The attention mechanism calculates attention weights to determine the contribution of each element. This way, the model can dynamically adjust weights based on the specific content of the input sequence, better capturing long-term dependencies and important patterns in the sequence.

However, transformer-based models need to address significant time and resource consumption issues. Many methods have been proposed to improve the performance and speed of attention-based models, reduce memory usage, and make them applicable to a wider range of time series prediction problems. For example, the original self-attention mechanism requires performing full connectivity computations across the entire input sequence, which can lead to prohibitively high computational costs for long sequences. As shown in Eq.~(\ref{ei1}), ProbSparse self-attention \cite{11} addresses this issue by introducing a sparsity-inducing mechanism, selectively interacting with only a subset of input positions based on probabilities, thereby reducing computational complexity while preserving model performance to some extent. We applied this mechanism to TempoScale, significantly reducing time costs and improving computational efficiency.
\begin{equation}
\label{ei1}
M\left(\mathbf{q}_i, \mathbf{K}\right)=\ln \sum_{j=1}^{L_K} e^{\frac{\mathbf{q}_i \mathbf{k}_j^{\top}}{\sqrt{d_k}}}-\frac{1}{L_K} \sum_{j=1}^{L_K} \frac{\mathbf{q}_i \mathbf{k}_j^{\top}}{\sqrt{d_k}}.
\end{equation}

In Eq.~(\ref{ei1}), \( M(\mathbf{q}_i, \mathbf{K}) \) represents the function computing the attention score, where \( \mathbf{q}_i \) is the query vector and \( \mathbf{K} \) is the set of key vectors. The query vector \( \mathbf{q}_i \) measures similarity between queries and keys, while \( \mathbf{K} \) represents the keys in the attention mechanism. Each \( \mathbf{k}_j \) is an element of the key vectors set with a length \( L_K \). The equation aims to compute the attention score efficiently, capturing the relationships between queries and keys.

Self-attention distillation \cite{11} aims to solve this problem by extracting the essential information captured by a large self-attention mechanism into a smaller one, while maintaining or even improving the model's performance. This method is implemented by integrating multiple pooling layers, as shown in Eq.~(\ref{ei3}):
\begin{equation}
\label{ei3}
\mathbf{X}_{j+1}^t=\operatorname{MaxPool}\left(\operatorname{ELU}\left(\operatorname{Conv1d}\left(\left[\mathbf{X}_j^t\right]_{\mathrm{AB}}\right)\right)\right).
\end{equation}

In Equation (\ref{ei3}), \(\mathbf{X}_{j+1}^t\) represents the feature representation of layer \(j+1\) at time step \(t\), which is obtained by applying a convolution operation followed by an Exponential Linear Unit (ELU) activation function, and then a max-pooling operation. The max-pooling operation selects the maximum value in each channel of the input tensor based on a specified window size, while the ELU activation function has a non-zero gradient in the negative region to alleviate the vanishing gradient problem. The convolution operation convolves the input tensor with a convolutional kernel to generate an output tensor, applied here to \(\left[\mathbf{X}_j^t\right]_{\mathrm{AB}}\) representing the feature representation of layer \(j\) at time step \(t\), where AB denotes a specific slice or subset selection. This equation describes the process of generating the feature representation of the next layer through convolution, activation, and max-pooling operations, where each operation can be controlled by adjusting parameters to influence the feature extraction process of the model.  Applying this mechanism to TempoScale enables the solution of prediction problems with longer sequences.

In traditional sequence generation tasks, each output is generated one at a time in a left-to-right manner, which can be slow and computationally expensive. The One Forward Decoder \cite{11} directly generates the entire long sequence without the need for individual generation, thus speeding up the generation process and reducing computational costs.

The aforementioned techniques contribute to enhancing the performance of transformer-based models while reducing time and memory overheads. TempoScale incorporates these techniques into its long-term prediction module to facilitate the execution of long-term forecasts.

\subsection{Obtaining the final results through a MLP}
In the end, the output of long-term and short-term prediction models, along with RC calculated in Algorithm~\ref{alg:ceemdan}, is fed into a MLP in TempoScale to obtain a final long-term time series prediction result for scaling. The MLP introduces non-linear transformations and higher-level feature representations, enabling more flexibility in capturing complex relationships between inputs and enhancing the model's understanding and generalization capabilities. In TempoScale, the MLP's input layer consists of 144 neurons, the output layer has 48 neurons, and there are 4 hidden layers with 192, 240, 240, and 192 neurons, respectively, all with ReLU activation functions.

Subsequently, the results are fed into the resource management system of a cloud cluster for resource auto-scaling. This step involves using the model's output to make resource management decisions, determining whether to allocate additional resources or remove excess resources. The overall goal of this process is to achieve more intelligent and efficient resource allocation, optimizing the performance of the entire system.

\section{Performance Evaluations}\label{sec:Performance}
In this section, we provide a detailed description of the dataset used and the experimental configurations. Additionally, we conducted experiments on the cluster to compare the performance of TempoScale with several state-of-the-art approaches. The results validate that TempoScale can be effectively applied to optimize cloud resource usage.

\subsection{Experimental setup}
TempoScale is mainly developed using Python 3.9. Resource scaling is performed every 15 seconds, and load prediction is conducted every 12 minutes. Load prediction utilizes data from the past 48 minutes to forecast the next 12 minutes. As shown in Table~\ref{tab:esDNN and Informer} and \cite{11}, the longer the predicted length, the greater the potential for improvement. Therefore, we'll take an intermediate value of 48 minutes for the prediction time length, and a prediction length of 12 minutes provides the system with a sufficient resource scheduling time \cite{spe2024}. All performance tests were conducted using a K8s cluster consisting of one master and two worker nodes. The operating system used was CentOS-7, with each node having 4 GB of memory and 4 CPU cores. The workload dataset, microservices demo application, and baseline methods used in the experiments are as follows:

\begin{figure}[H]
	\centerline{\includegraphics[width=\linewidth]{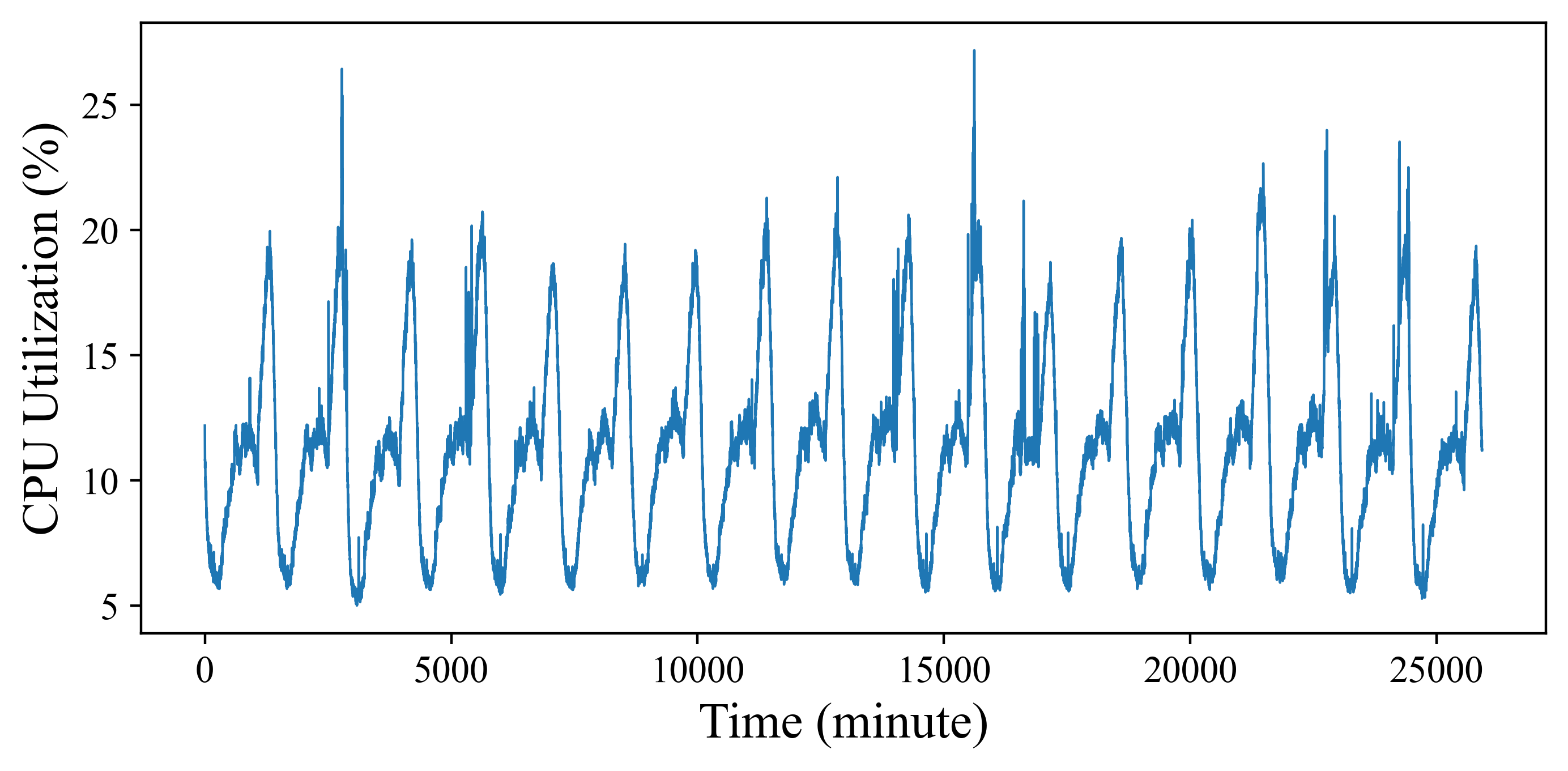}}
	\caption{CPU Utilization of \textit{MS\_10489} Over Time.}\label{fig:MS}
\end{figure}

\subsubsection{Workload dataset}
We used the dataset from the Alibaba Cluster\footnote[4]{https://github.com/alibaba/clusterdata/tree/master/cluster-trace-microservices-v2022}, which was collected from Alibaba production clusters consisting of over ten thousand bare-metal nodes over a period of 13 days in 2022. One of the datasets, named \textit{MSResource}, records CPU and memory utilization of over 470,000 containers for more than 28,000 microservices in the same production cluster. It includes attributes such as \textit{timestamp}, \textit{msname}, \textit{msinstanceid}, \textit{nodeid}, \textit{cpu\_utilization}, and \textit{memory\_utilization}. This dataset accurately represents the workload characteristics of current large-scale cloud clusters. We utilized this dataset as input for workload simulation to evaluate the performance and reliability of applications or systems under various workload conditions. For experimental evaluation, we select a microservice named \textit{MS\_10489}, illustrating the variation in its resource utilization rates in Fig.~\ref{fig:MS}.

\subsubsection{Microservices demo application}
Sock Shop\footnote[5]{https://microservices-demo.github.io/} is a microservice application commonly used for testing purposes. It is an open-source demo application designed to demonstrate best practices in developing cloud native applications. Sock Shop simulates an online shopping platform and consists of eight microservices, each serving a specific function such as shopping carts, payments, and inventory.

\subsubsection{Baseline methods}
The three baseline methods used in our experiments are state-of-the-art and representative methods of the three categories discussed in Section~\ref{sec:Related Work}.

\begin{enumerate}[]
	\item \textbf{ARIMA} \cite{box2015time}: It effectively captures trends and seasonality in time series data. Due to its simplicity and widespread application, ARIMA is often used as a standard for comparing the performance of other time series forecasting models.
	\item \textbf{esDNN} \cite{2}: This is an optimized method based on GRU. It is designed to be simple, with fewer parameters, easy to train, and computationally efficient. It serves as an ideal benchmark for capturing sequential patterns in various tasks..
	\item \textbf{Informer} \cite{11}: This is an improved method based on Transformer. Due to its outstanding performance in sequence tasks and the effectiveness of its self-attention mechanism in handling long-term dependencies, it is the preferred benchmark method for many sequence data processing tasks.
\end{enumerate}

\subsection{Profiling}\label{Profiling}
\begin{figure}
	\centerline{\includegraphics[width=\linewidth]{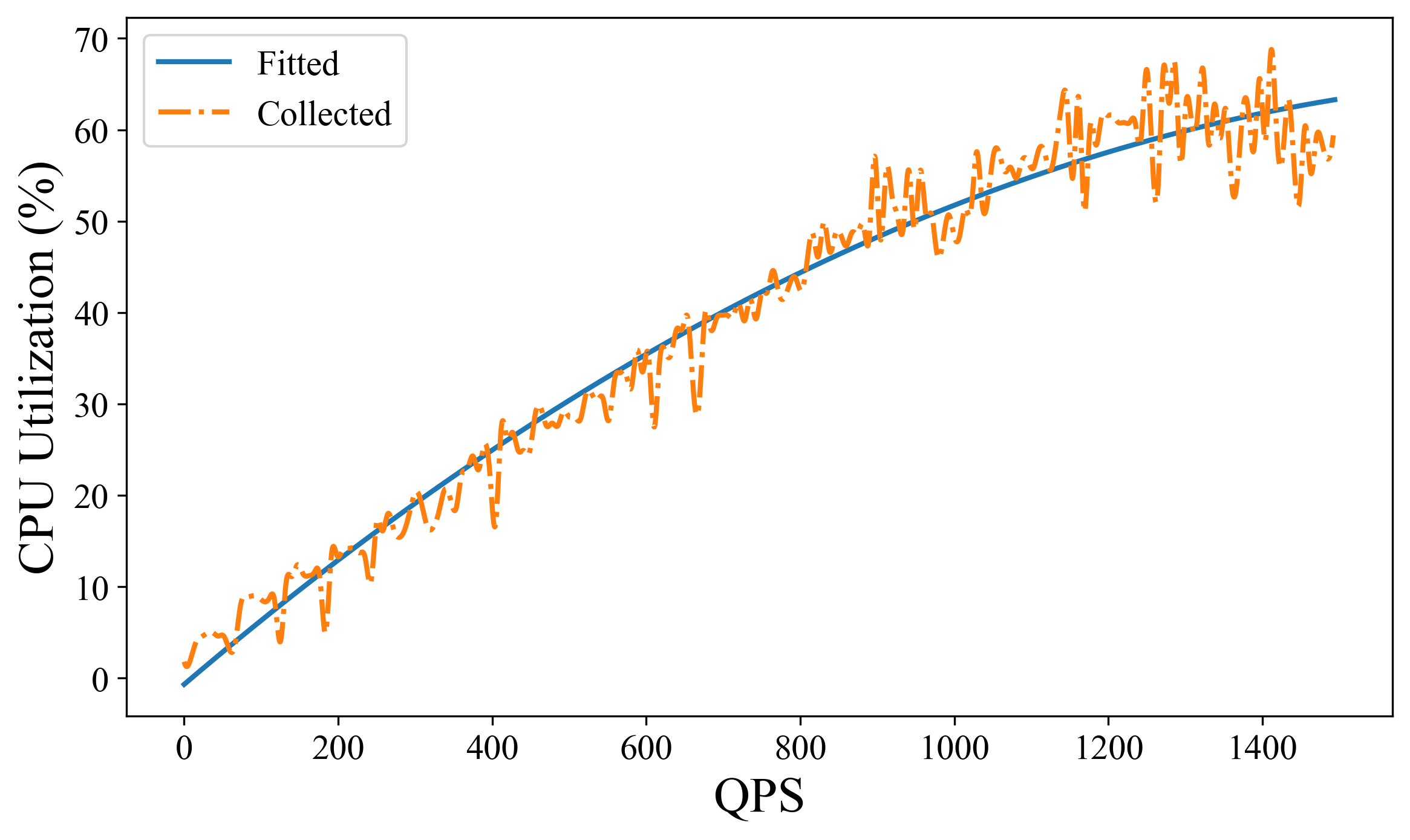}}
	\caption{The Profiling Between QPS and CPU Utilization.}
	\label{fig:sock-shop}
\end{figure}

Due to the varying relationship between CPU utilization and Queries Per Second (QPS) on each machine, which depends on factors such as hardware configuration, load characteristics, and the running software system, it is necessary to establish the profile between CPU utilization and QPS before starting the experiment. This helps determine the expected CPU utilization levels at different QPS levels, ensuring the accuracy and comparability of the experimental results. The profiling results are shown in Fig.~\ref{fig:sock-shop}, the experimental results are similar to those of previous work \cite{PerformanceModeling}.

% \begin{figure*}
% 	\centering
% 	\subfigure[ARIMA.]{\label{fig:subfig:a}
% 		\includegraphics[width=0.49\textwidth]{arima.png}}
% 	\subfigure[esDNN.]{\label{fig:subfig:b}
% 		\includegraphics[width=0.49\textwidth]{esdnn.png}}
% 	\subfigure[Informer.]{\label{fig:subfig:c}
% 		\includegraphics[width=0.49\textwidth]{informer.png}}
% 	\subfigure[TempoScale.]{\label{fig:subfig:d}
% 		\includegraphics[width=0.49\textwidth]{temposcale.png}}
% 	\caption{Runtime Performance Analysis.}
% 	\label{fig01}
% \end{figure*}

\subsection{Predictive Evaluation}

Table~\ref{tab:result} presents the forecast results of ARIMA, esDNN, Informer, and TempoScale on Alibaba's 2022 trace data, evaluated using the Mean Square Error (MSE), coefficient of determination (R$^2$), and Mean Absolute Percentage Error (MAPE), The equations are as follows:
\begin{equation}
\text{MSE} = \frac{1}{n} \sum_{i=1}^{n} (y_i - \hat{y}_i)^2,
\end{equation}
\begin{equation}
\text{MAPE} = \frac{1}{n} \sum_{i=1}^{n} \left| \frac{y_i - \hat{y}_i}{y_i} \right| \times 100,
\end{equation}
\begin{equation}
\text{R}^2 = 1 - \frac{\sum_{i=1}^{n} (y_i - \hat{y}_i)^2}{\sum_{i=1}^{n} (y_i - \bar{y})^2},
\end{equation}
where $n$ is the sample size, $y_i$ represents the $i$th observed value, $\hat{y}_i$ represents the model's predicted value for the $i$th observation, $\bar{y}$ represents the mean of the observed values. 

\begin{table}
\caption{Comparison of Prediction Results from Different Methods.}
\label{tab:result}
\centering
\resizebox{\linewidth}{!}{%
\begin{tblr}{
  width = \linewidth,
  colspec = {Q[156]Q[187]Q[202]Q[187]Q[187]},
  cells = {c},
  hline{1,5} = {-}{0.08em},
  hline{2} = {-}{0.05em},
}
Method & ARIMA      & esDNN     & Informer  & TempoScale\\
MSE    & 0.000099~  & 0.000085~ & 0.000073~ & \textbf{0.000069}\\
MAPE   & 0.049752~  & 0.047577~ & 0.048178~ & \textbf{0.044682}\\
R$^2$      & 0.359305~  & \textbf{0.400834} & 0.358701~ & 0.365229~
\end{tblr}
}
\end{table}

As shown in Table~\ref{tab:result}, we have highlighted the best value for each metric. Assessment metrics are calculated on a per-prediction-period basis, and the displayed results have all been subjected to inverse normalization. The data demonstrates that our proposed method, TempoScale, outperformed others in terms of both MSE and MAPE. Specifically, in terms of MSE, TempoScale outperforms ARIMA by 30.43\%, esDNN by 18.78\%, and Informer by 5.80\%. In terms of MAPE, TempoScale outperforms ARIMA by 10.19\%, esDNN by 6.08\%, and Informer by 7.26\%. Although it did not yield the highest result in terms of R$^2$, it remained at a satisfactory level. The above results suggest that TempoScale exhibits higher accuracy and reliability in forecasting Alibaba Cloud workload data.

\begin{figure*}
	\centering
	\subfigure[The First Time Step of the Prediction Interval.]{\label{fig:start}
		\includegraphics[width=0.485\textwidth]{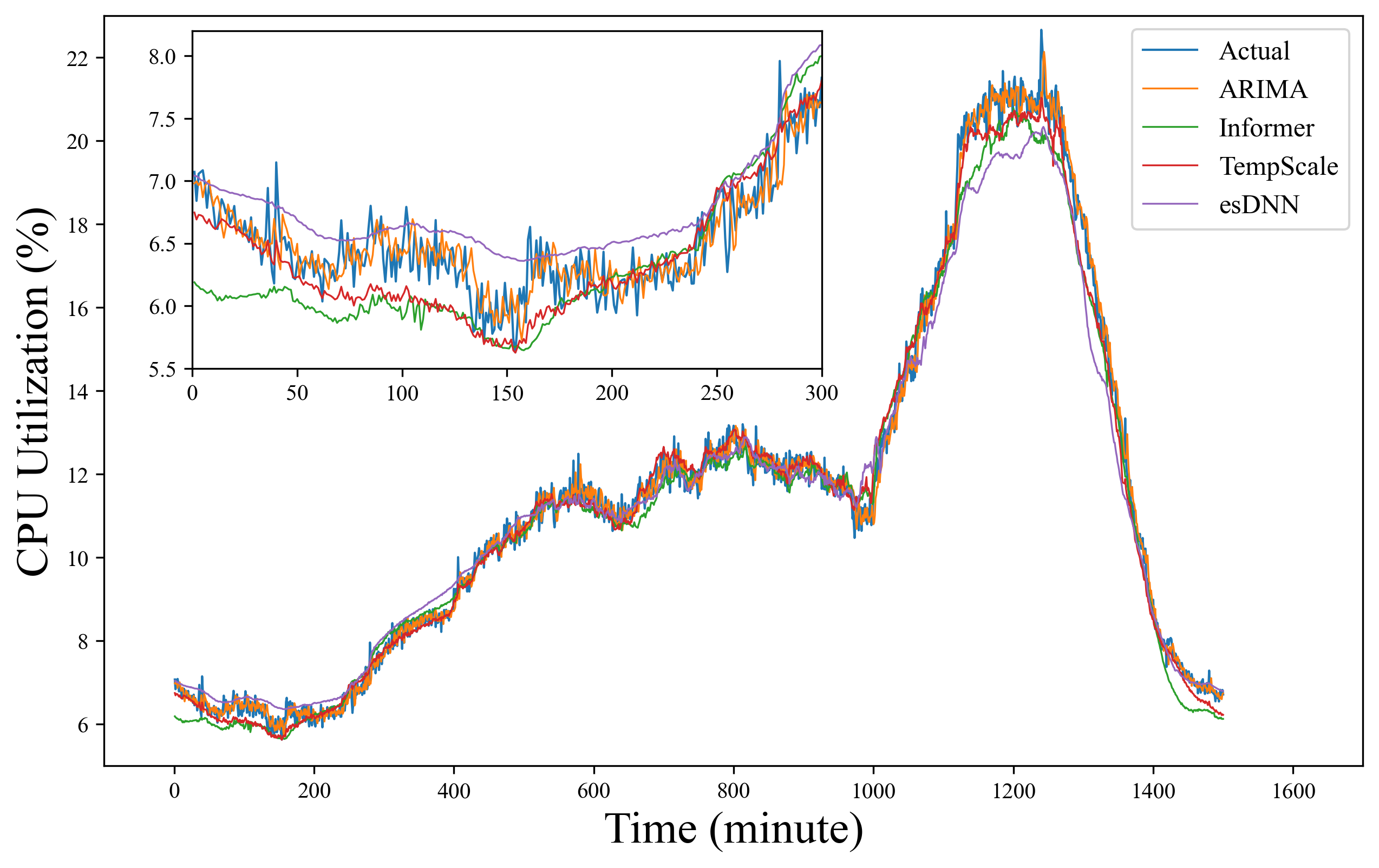}}
	\subfigure[The Last Time Step of the Prediction Interval.]{\label{fig:end}
		\includegraphics[width=0.485\textwidth]{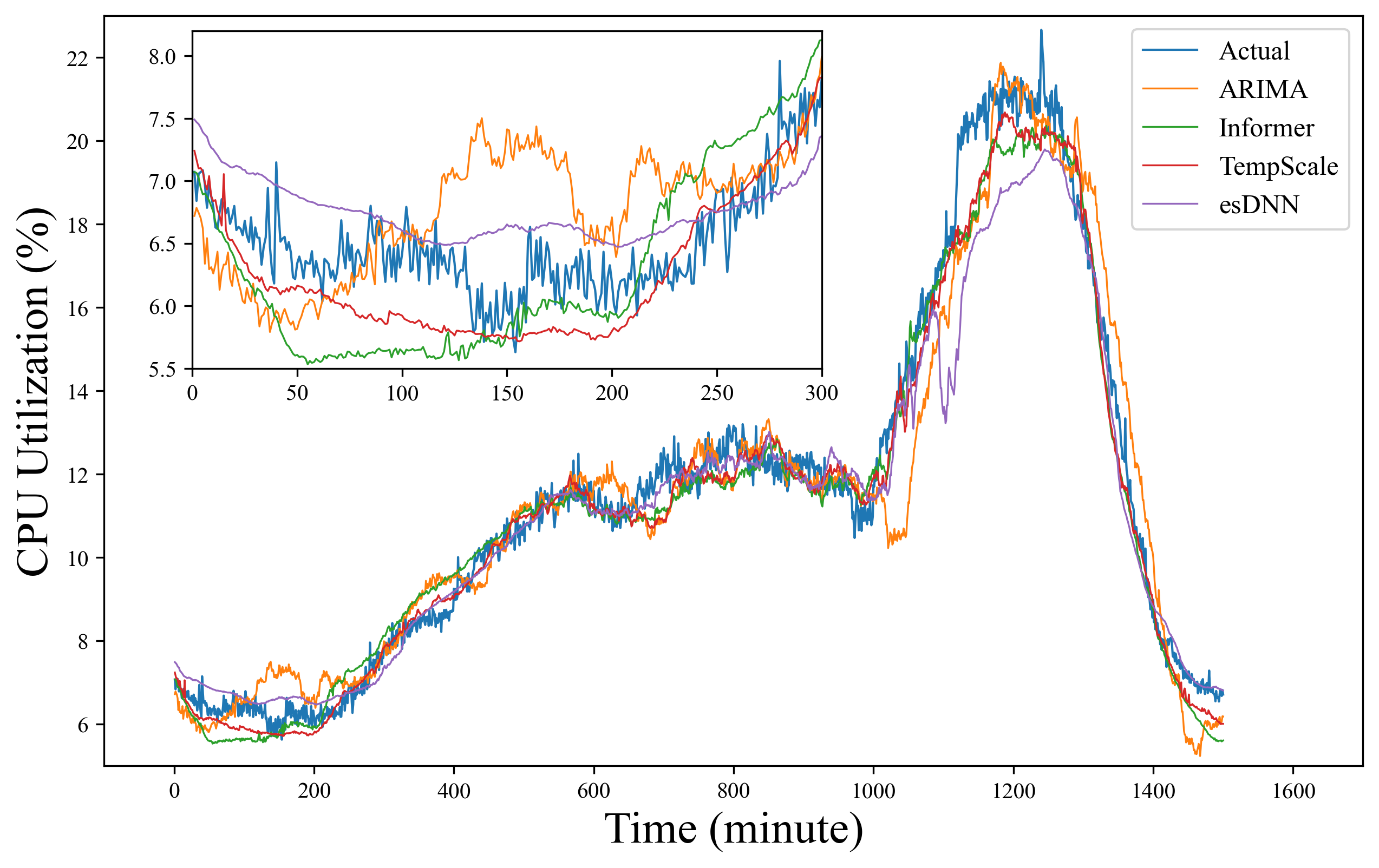}}
	\caption{Comparison Between Predicted and Actual Values Based on Alibaba Dataset.}
	\label{fig:start and end}
\end{figure*}

To investigate the impact of forecast length on accuracy, we selected a subset of forecast results (1500 data points, with each data point representing one minute) for visualization. Fig.~\ref{fig:start} illustrates the forecast results for the first time slice within the forecasting period, demonstrating the outstanding performance of ARIMA, with its predictions almost perfectly aligning with the actual data. Fig.~\ref{fig:end} demonstrates the forecast results for the last time slice within the forecasting period. Upon inspection in the zoomed-in figure, it is evident that Informer and esDNN either underestimated or overestimated the actual data, whereas TempoScale was able to predict the actual data more accurately.

\begin{figure*}
	\centering
	\subfigure[MSE.]{\label{fig:MSE}
		\includegraphics[width=0.332\textwidth]{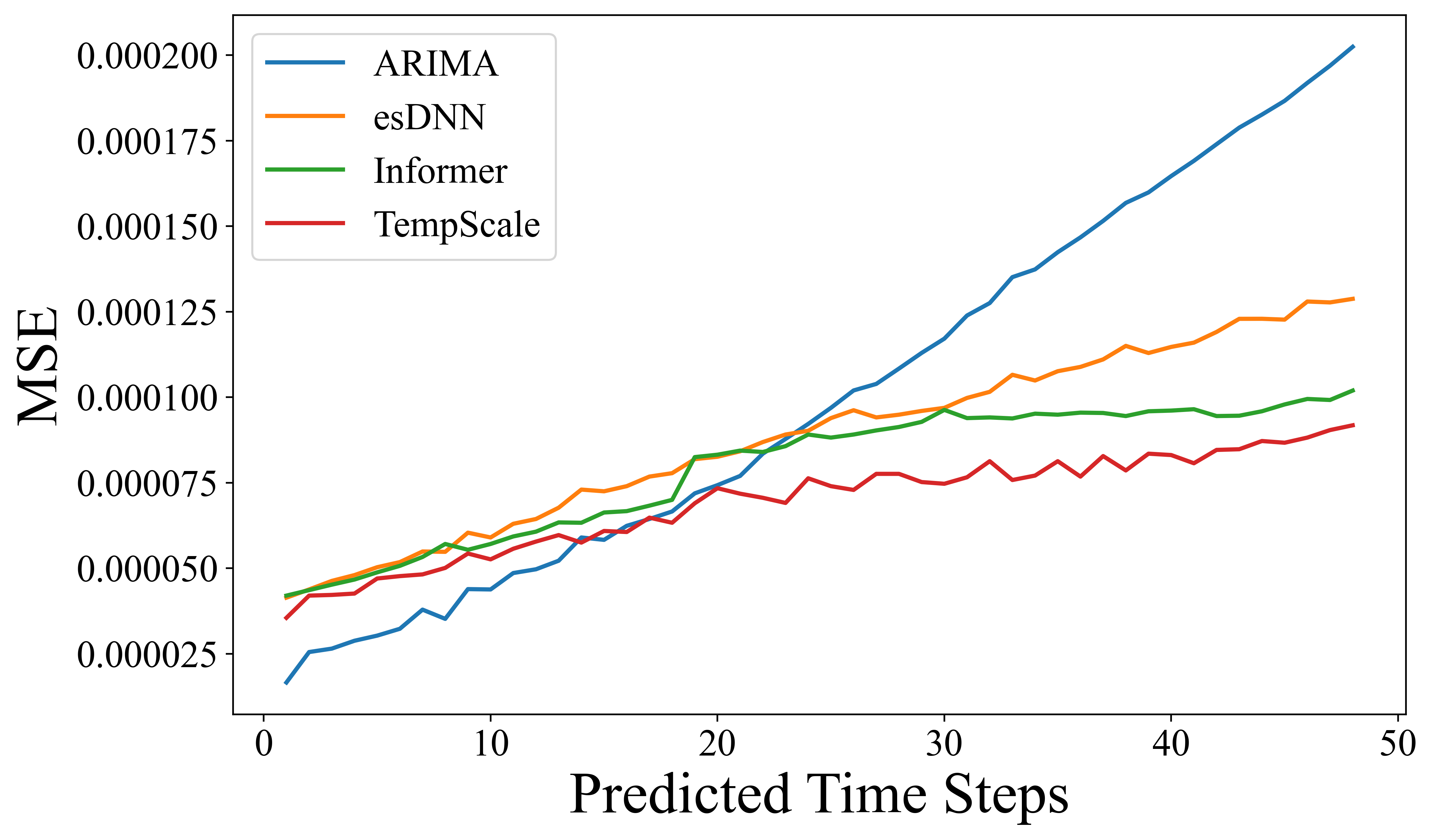}}
	\subfigure[MAPE.]{\label{fig:MAPE}
		\includegraphics[width=0.317\textwidth]{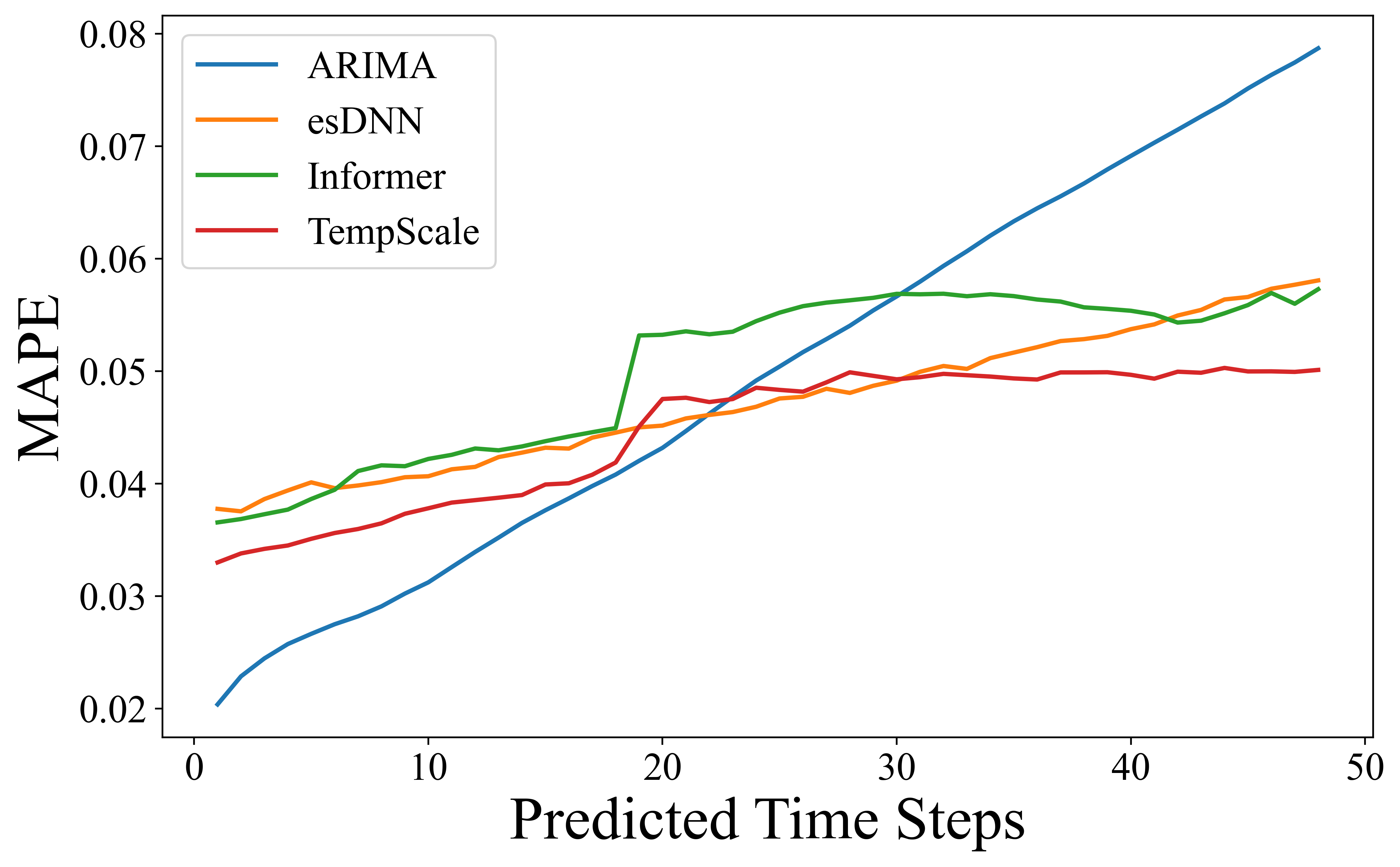}}
	\subfigure[R$^2$.]{\label{fig:R2}
		\includegraphics[width=0.317\textwidth]{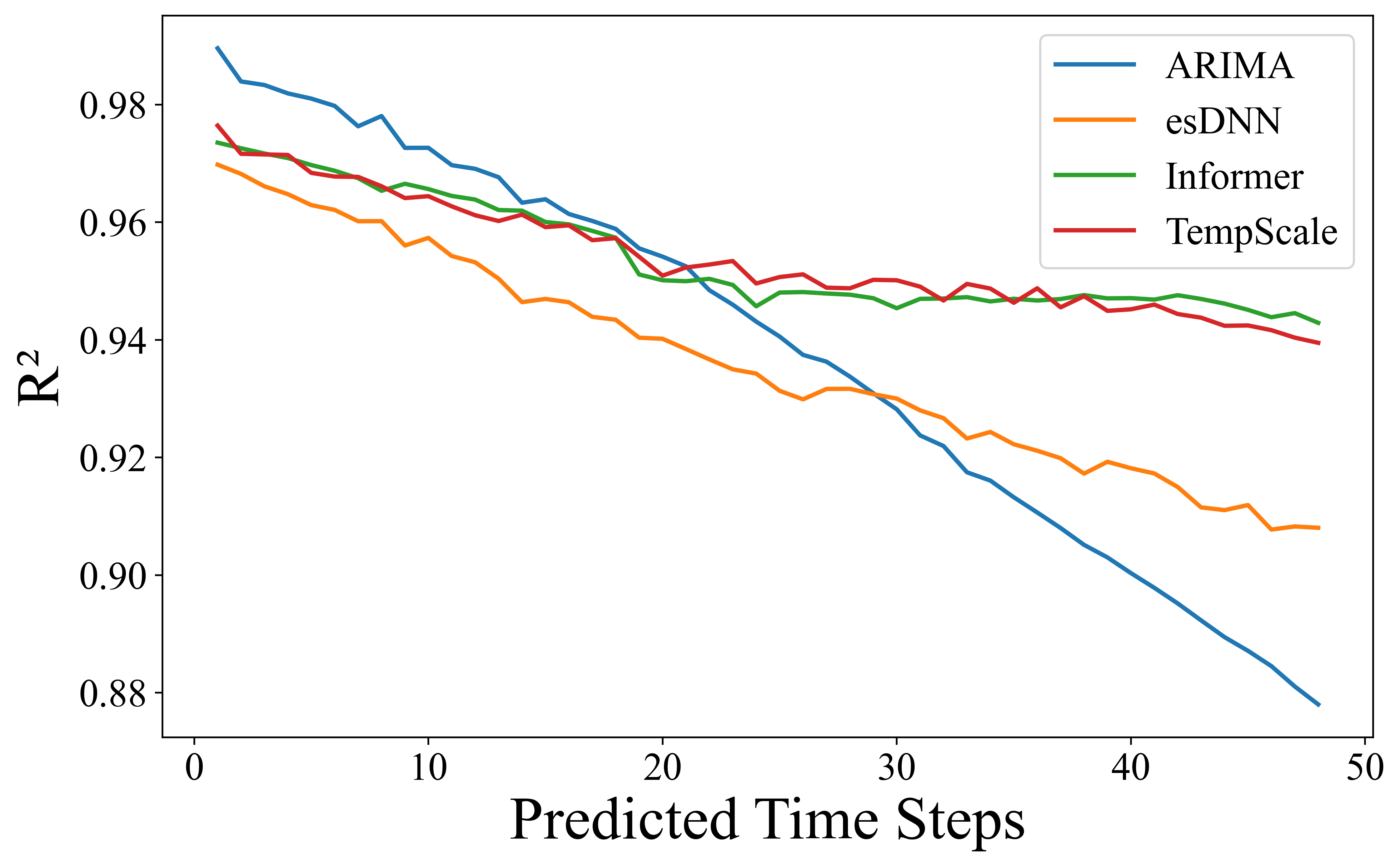}}
	\caption{Comparison of Prediction Performance Under Different Time Steps Based on Alibaba Dataset.}
	\label{fig:result}
\end{figure*}

In order to study the impact of time steps on prediction accuracy, we conducted statistical calculations to analyze the variations in performance metrics across different time steps. It is evident from Fig.~\ref{fig:result} that the performance of ARIMA deteriorates almost linearly with the increase in the length of the time interval, as indicated by MSE, MAPE, and R$^2$ in Fig.~\ref{fig:MSE}, Fig.~\ref{fig:MAPE}, and Fig.~\ref{fig:R2}. The reason lies in that ARIMA may overly rely on the most recent historical data in time series forecasting tasks, treating this segment as the prediction result while overlooking earlier historical data variations. This behavior could lead to a strong correlation between the prediction results and the most recent historical data segment. Conversely, the degradation rate of performance for esDNN and Informer is relatively slower, this 
results from that they are able to effectively model long-term dependencies, capturing long-term correlations in time series data through techniques such as self-attention mechanisms, thereby maintaining good performance even with longer time steps. TempoScale combines the advantages of both, thereby exhibits nearly the slowest deterioration in performance, demonstrating its outstanding performance in long-term forecasting.

\subsection{Workload Prediction with Auto-Scaling Evaluation}

The benefits of auto-scaling lie in its ability to effectively manage costs and improve system performance and availability. By automatically adjusting resource usage based on actual workload, this technology avoids resource waste and shortages, thus saving costs. Additionally, it ensures system performance during high workloads and reduces resource usage during low workloads, enabling efficient system operation. 

Therefore, to further demonstrate the capability of the proposed method and develop an efficient auto-scaling approach, we integrate methods including ARIMA, esDNN, Informer and TempoScale into the prototype system based on K8s developed by us are conducted to evaluated.

We first simulate workload variations in a real cluster environment using Locust\footnote[6]{https://locust.io/}, based on the results of the profiled data in Section~\ref{Profiling}. Then, we utilize the elastic scaling system integrated with workload prediction methods to evaluate the effectiveness of this method in cost management and performance improvement. We focus on vertical scaling of containers, with predictions operating on a 12-minute cycle. Vertical scaling involves adjusting the resource configuration of individual instances, such as increasing the CPU quota or memory limit of containers in a containerized environment. The amount of scaling operations is based on the predictions of CPU usage. Our goal is to enhance system performance by minimizing response time and avoiding SLO violations while keeping the total resource budget constant. Here, the resource budget refers to the cumulative product of resource supply within each time unit, represented as $\int R_t \, dt$, where $R_t$ is the resource provided at time $t$.

\begin{figure}
	\centerline{\includegraphics[width=\linewidth]{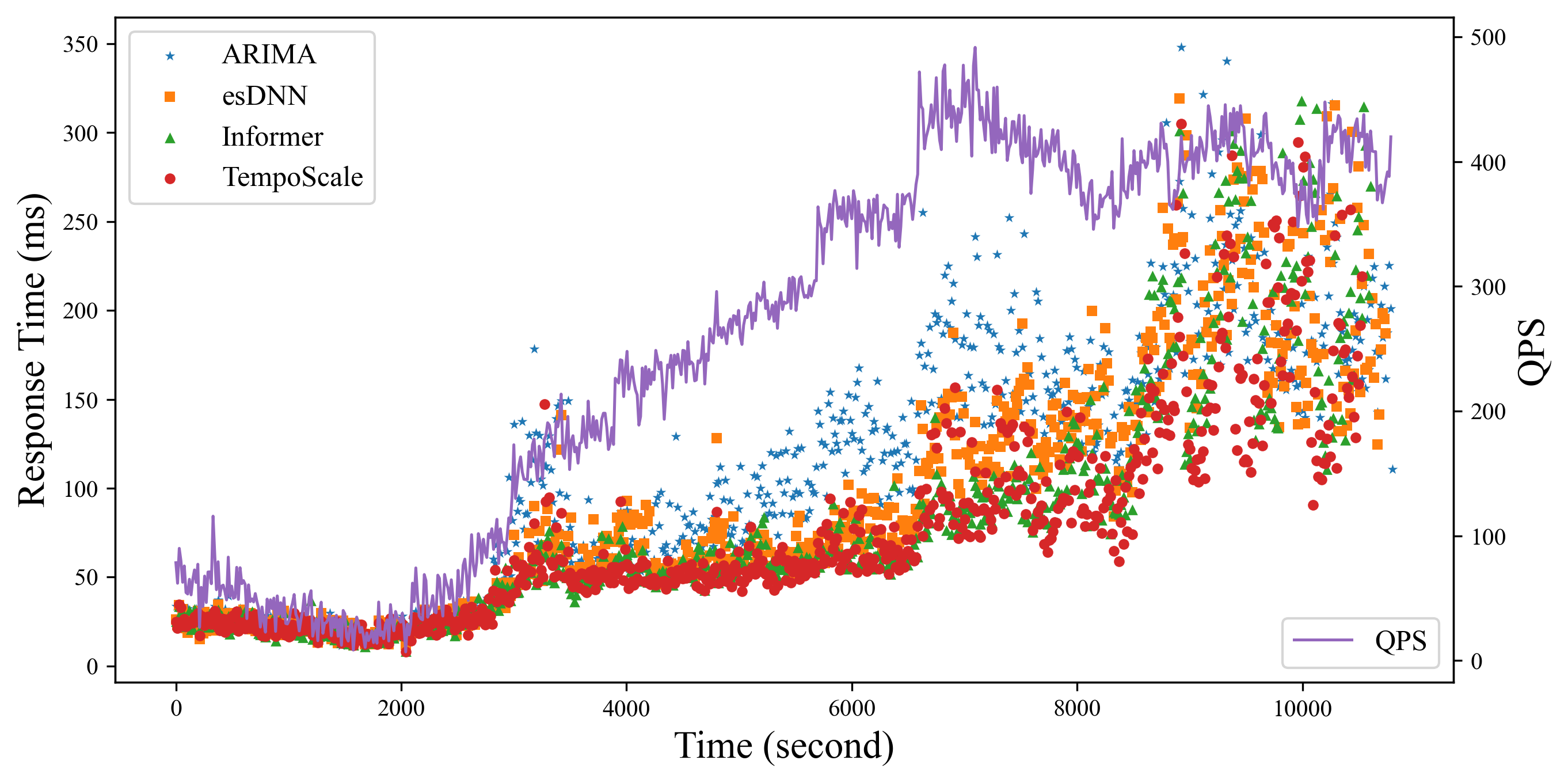}}
	\caption{Comparison of Average Response Times during Runtime.}
	\label{fig:responsetime}
\end{figure}

\begin{figure}
	\centerline{\includegraphics[width=\linewidth]{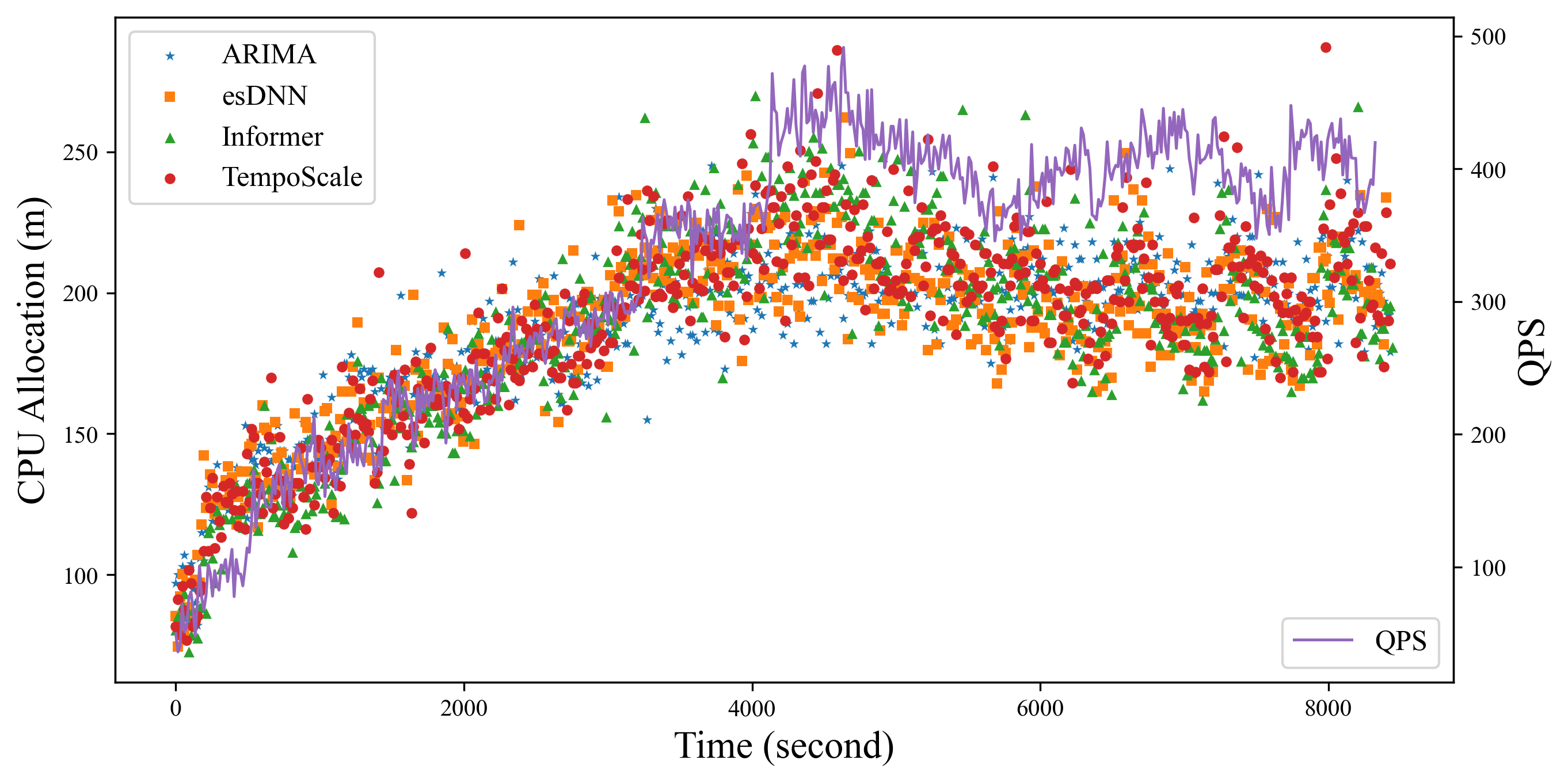}}
	\caption{Comparison of CPU Allocation during Runtime.}
	\label{fig:allocation}
\end{figure}

In the experiments, detailed average response time is shown in Fig.~\ref{fig:responsetime}, while CPU allocation is depicted in Fig.~\ref{fig:allocation}, with units in milli-cores (m), this unit of measurement is commonly used to specify the amount of CPU resources that an application or container can utilize in cloud computing and containerized environments. The figures include CPU allocation, average response time, and workload size for each method during runtime. Additionally, QPS values are plotted on the secondary axis of each figure to facilitate direct comparison. From the figures, it can be observed that during the initial stable workload phase, all four methods maintain system stability with relatively low average response times. However, in the long-term stages, workload spikes and variations lead to varying degrees of response time increases. Notably, TempoScale demonstrates lower performance impact compared to other methods, while ARIMA exhibits the poorest performance. This aligns with the results from the previous section on workload prediction experiments.

\begin{table}
\centering
\caption{Comparison of Response Time, SLO Violation Rate and Resource Usage.}
\label{tab:experimental data1}
\resizebox{\linewidth}{!}{%
\begin{tblr}{
  cells = {c},
  hline{1,9} = {-}{0.08em},
  hline{2} = {-}{0.05em},
}
Method & ARIMA & esDNN & Informer & TempoScale\\
Average Response Time (ms) & 110.52 & 90.38 & 80.59 & \textbf{76.09}\\
99th Percentile Response Time (ms) & 378.85 & 327.82 & 319.50 & \textbf{314.91}\\
Maximum Response Time (ms) & 471.07 & 409.75 & 399.45 & \textbf{396.02}\\
SLO (250 ms) Violation (\%) & 2.64 & 2.50 & 3.47 & \textbf{1.39}\\
SLO (200 ms) Violation (\%) & 10.69 & 8.06 & 8.61 & \textbf{4.58}\\
CPU Budget (m$\cdot$s) & 119938.00 & 119285.52 & 120973.88 & 119912.67\\
CPU Usage (m$\cdot$s) & 92466.70 & 97927.95 & 103548.11 & \textbf{108095.69}
\end{tblr}
}
\end{table}

Finally, we compare the response time and SLO violation rates as shown in Table~\ref{tab:experimental data1}, CPU budget and usage are measured in m$\cdot$s, representing the product of time and resource quantity, the experiments are conducted under roughly the same resource budget to compare resource usage. The average response time of the TempoScale method is 76.09 ms, achieving best performance. It outperforms ARIMA by 31.15\%, esDNN by 15.81\%, and Informer by 5.58\%. As for the SLO violation rates, users' acceptance may vary depending on specific application scenarios and business requirements. Generally, most users expect fast response times and high-performance services, so shorter SLOs (e.g., a few hundred milliseconds or shorter) are typically considered ideal. Here, we set two SLO targets, 200 ms and 250 ms, reflecting these expectations. When the SLO is set to 200 ms, the violation rate for TempoScale is 4.58\%, which is 6.11\% lower than ARIMA, 3.48\% lower than esDNN, and 4.03\% lower than Informer.

Based on the above results, it can be concluded that compared to more primitive forecasting algorithms such as ARIMA, TempoScale can improve performance by over 30\%. TempoScale can also improve performance by 5-10\% over novel and innovative algorithms such as Informer and esDNN, which have been proposed recently. Moreover, the optimization approach proposed by TempoScale will also contribute to efficiency enhancement for enterprises, promoting the development of various scenarios (e.g. auto-scaling) in cloud.

\section{Conclusions and Future Work}\label{sec:Conclusions}
In order to address the inherent dynamics of clusters and the variability of workloads, we have proposed an innovative solution called TempoScale. It is designed to better capture the correlations in time series data, enabling more intelligent and adaptive elastic scaling decisions. TempoScale utilizes long-term trend analysis to reveal the changes in workload and resource demands, supporting proactive resource allocation over extended periods. Additionally, it employs short-term volatility analysis to examine variations in workload and resource demands, facilitating real-time scheduling and rapid responsiveness. We conducted experiments on top of K8s with realistic data from Alibaba, and the results demonstrate the feasibility of our proposed method. Our approach not only enhances system performance and stability but also effectively reduces resource costs, promoting the sustainable development of cloud computing across various industries. However, the framework of TempoScale is built on individual microservices without fully considering the invocation dependency graph \cite{b14} and emergency measures in cases of inaccurate predictions. In future work, we plan to address these aspects to enhance TempoScale's capability in handling microservices with complex dependency graphs and improving robustness in special situations, and exploring additional integration possibilities with cloud management platforms.

\section*{SOFTWARE AVAILABILITY}
The codes have been open-sourced to \url{https://github.com/lifwen/TempoScale} for research usage.

\bibliographystyle{IEEEtran}
\bibliography{bibliography.bib}

\vspace{12pt}

\end{document}